# Observation of a d-wave nodal liquid in highly underdoped $Bi_2Sr_2CaCu_2O_{8+\delta}$

U. Chatterjee[1,2], M. Shi[3], D. Ai[1,2], J. Zhao[1,2], A. Kanigel[4], S. Rosenkranz[2], H. Raffy[5], Z. Z. Li[5], K. Kadowaki[6], D. G. Hinks[2], Z. J. Xu[7], J. S. Wen[7], G. Gu[7], C. T. Lin[8], H. Claus[2], M. R. Norman[2], M. Randeria[9], and J. C. Campuzano[1,2*]

[1]Department of Physics, University of Illinois at Chicago, Chicago, IL 60607
[2]Materials Science Division, Argonne National Laboratory, Argonne, IL 60439
[3]Swiss Light Source, PSI, CH-5232 Villigen, Switzerland
[4]Department of Physics, Technion, Haifa 32000, Israel
[5]Laboratoire de Physique des Solides, Université Paris-Sud CNRS-UMR 8502, 91405 Orsay Cedex, France
[6]Institute of Materials Science, University of Tsukuba, Ibaraki 305-3573, Japan
[7]Physics Department, Brookhaven National Laboratory, P.O. Box 5000, Upton, NY 11973
[8]Max Planck Institute for Solid State Research, D-70569 Stuttgart, Germany
[9]Department of Physics, The Ohio State University, Columbus, OH 43210

We use angle resolved photoemission spectroscopy to probe the electronic excitations of the non-superconducting state that exists between the antiferromagnetic Mott insulator at zero doping and the superconducting state at larger dopings in $Bi_2Sr_2CaCu_2O_{8+\delta}$. We find that this state is a nodal liquid whose excitation gap becomes zero only at points in momentum space. Despite exhibiting a resistivity characteristic of an insulator and the absence of coherent quasiparticle peaks, this material has the same gap structure as the d-wave superconductor. We observe a smooth evolution of the spectrum across the insulator-to-superconductor transition, which suggests that high temperature superconductivity emerges when quantum phase coherence is established in a non-superconducting nodal liquid.

High temperature superconductivity in the cuprates occurs by doping a Mott insulator whose antiferromagnetic ground state and low-energy excitations are well understood[1]. By adding carriers, the parent insulator turns into a superconductor for dopings that exceed 0.05 holes per $CuO_2$ plane. The d-wave nature of the superconducting ground state[2] and its low-lying excitations are also well understood. In between these two phases lies an electronic ground state whose nature is poorly understood. As the temperature is raised, this intermediate "pseudogap" state occupies a larger and larger region of the phase diagram (Fig. 1a). It is from this phase that the superconducting state emerges for all but the most highly doped samples. Consequently, the nature of this phase holds the key to the origin of high temperature superconductivity.

While the electronic excitations in the high temperature pseudogap region have been studied extensively, there is little spectroscopic data at low temperatures, as there is only a very narrow window of dopings where neither superconducting nor antiferromagnetic order occurs. Here we present angle-resolved photoemission spectroscopy (ARPES) data on single crystals and thin films[3] whose doping levels range all the way from the insulator to the over-doped superconductor. We focus in particular on non-superconducting thin films, just to the left of the superconducting $T_c$ dome (see Fig. 1a), with an estimated hole doping $\sim 0.04$[3]. These samples have an insulating upturn in resistance $R(T)$ with decreasing temperature shown in Fig. 1b, that is well described by 2D-variable range hopping[4,5] (See Supplemental section). We have measured the diamagnetic susceptibility down to 1.5 K, and found no trace of superconductivity, with a sensitivity of 1% of the volume fraction. We compare the results on the non-superconducting samples with data on superconducting ones.

In Fig. 1d we show the energy distribution curves (EDCs, spectra at constant momentum **k** as a function of binding energy $\omega$). Despite the low temperature, no sharp, coherent features are discernable in the spectra. This is not surprising, since earlier work had found a strong suppression of coherent spectral weight in the superconducting state with underdoping[6]. In contrast, the momentum distribution curves (MDCs) in Fig. 1e at $\omega = 0$ show clearly visible peaks. Thus the excitations are much better defined in **k**-space than they are in energy, and are sharper near the zone diagonal than near its boundary (the Brillouin zone is shown in Fig. 1c). Remarkably, despite the insulating-like nature of the resistivity, the MDC peaks indicate a locus of minimum energy excitations similar to that of the superconductors, clearly visible in the ARPES intensity map in Fig. 1c at $\omega = 0$. These "Fermi momentum" ($\mathbf{k}_F$) values were in fact used to generate the EDCs in Fig. 1d. In Fig. 1f we show the ARPES intensity as a function of energy and **k** for another sample at the same doping, which shows that there is a well-defined dispersion despite the incoherent nature of the EDCs.

To better understand the electronic excitations in the non-superconducting sample in Fig. 1d, we plot its raw EDCs at $\mathbf{k}_F$, symmetrized to remove the effects of the Fermi function[7], and compare them with superconducting state spectra (Figs. 2e, f, and g) at various doping indicated in Fig. 2a. In each panel d through g, the top curve corresponds to $k_F$ on the zone boundary ($\theta = 0°$), while the lowest curve to $k_F$ on the zone diagonal ($\theta = 45°$), with the Fermi surface angle $\theta$ increasing from top to bottom.

In the non-superconducting sample (Fig. 2d) we see a highly anisotropic energy gap which decreases monotonically from a maximal value at $\theta = 0°$, to zero at $\theta = 45°$ Even though there are no sharp coherence peaks at any angle, there is a clearly discernable *low-energy* gap. This is the pseudogap at low temperature (16 K) in the non-superconducting sample. With increasing doping we move from the non-superconductor (Fig. 2d) to the highly underdoped superconductor (Fig. 2e), whose EDCs look qualitatively similar to those in Fig. 2d, except for the appearance of observable coherent quasiparticle peaks at the gap edge at each angle on the Fermi surface. These peaks grow in strength with increasing doping (Fig. 2f and g). We note that the energy gap evolves smoothly going from the insulator (Fig. 2d) to the optimally doped superconductor (Figs. 2e to g) as the doping increases.

Much of the intensity of the EDCs in Fig. 2d can be traced to the existence of a large "background" that is present for all **k**'s in the zone. The origin of this background is not entirely clear, though it can be readily identified as the ARPES spectra for unoccupied states (k much beyond $k_F$), or extracted from the 'flat' (**k**-independent) part of the MDCs[8], and is plotted in Fig. 2b. We note that following the signal, the intensity of the background also continuously decreases as the doping decreases (See supplemental information). If this background is subtracted from the data in Fig. 1d, we obtain the symmetrized results shown in Fig. 2c, which further emphasize the presence of a low energy pseudogap ($\theta = 0°$) and the node ($\theta = 45°$).

In Fig. 3 we show the angular anisotropy of the spectral gap for all our thin film and single crystal samples. The superconducting samples were measured at temperatures between 16K and 40K, well below their $T_c$; the non-superconducting sample was measured at 16 K. The energy gap for superconducting samples was determined from half the spacing between coherence peaks in symmetrized spectra at their corresponding $k_F$. For the non-superconducting sample, we determine the low energy gap from the raw data (as indicated by the intersection of red straight lines in Fig. 2d), from the background subtracted data (Fig. 2c), and also from lineshape fits. All three methods lead to the same gap estimates within error bars.

Normalizing the gaps to their maximal value at $\theta = 0°$, we find the surprising result (Fig. 3a) that they follow a simple $|\cos(2\theta)|$ behavior for *all* samples, superconducting and non-superconducting. This '*d*-wave' gap is universal in nature, and does not distinguish between the *d*-wave superconductor and the low temperature pseudogap phase. The maximal energy gap values, plotted in Fig. 3b, monotonically increase with underdoping as found earlier[6], although they might decrease near the superconductor-insulator boundary.

Our results confirm an earlier extrapolation[9], based on ARPES measurements above $T_c$ for underdoped superconducting samples, that the low temperature pseudogap phase should be characterized by a node along the zone diagonal. They are also consistent with thermal conductivity ($\kappa$) measurements[10] in highly underdoped $YBa_2Cu_3O_{6+\delta}$, which show that the low temperature $\kappa/T$ of the insulating phase proximate to the superconducting dome is the same as that in the $d$-wave superconducting phase, where it is dominated by nodal excitations.

Many experiments report a node in the superconducting state but a gap that deviates from the simple $|\cos(2\theta)|$ form with underdoping. This behavior is attributed to two different order parameters[11], with an energy gap in the antinodal region (near $\theta = 0°$) larger than what would be inferred by extrapolating the gap from the nodal region ($\theta = 45°$). Our own work[12] a decade ago found evidence for a flattening of the gap around the nodes, but at that time the detectors had at least an order of magnitude lower **k**-resolution and sparser angular sampling compared to the present study. Recent ARPES studies[13-15] and scanning tunneling microscopy[16,17] have also reported 'two gap' behavior.

The present results are not consistent with a two gap picture. Several other experiments also find evidence for a simple $d$-wave gap. These include thermal transport data[18], where the nodal gap slope extrapolated to the antinode was consistent with the maximum gap, and ARPES data on underdoped $La_{2-x}Sr_xCuO_4$[19], 1/8 doped $La_{2-x}Ba_xCuO_4$[20], and Bi2201[21,22]. Why some experiments and/or samples show 'two-gap' behavior while others show a simple $d$-wave gap is not presently understood. We must emphasize here that we observe coherent quasiparticle peaks at the gap edge at all $\mathbf{k}_F$ for all superconducting samples down to the lowest $T_c$'s. If, however, quasiparticle peaks were absent near the antinodes, one would erroneously estimate much larger gap values in the vicinity of these **k**-points.

Another important question in highly underdoped samples is the possible existence of hole pockets, reported in a recent ARPES measurement[24]. We have not found any evidence for such pockets. Our MDCs always trace out a large underlying Fermi surface as in Fig. 1c.

This brings us to the implications of our main result, the observation of a d-wave like gap in a non-superconducting state that persists through the insulator-to-superconductor transition. One possibility is that our insulating sample is highly inhomogeneous and has a small fraction of superconducting regions that dominate the low-energy signal, while the less doped insulating regions produce the large spectroscopic background and dominate the transport. This inhomogeneity should be intrinsic, and not a surface phenomenon, since we have found very similar results in the superconducting samples for thin films and single crystals of $Bi_2Sr_2CaCu_2O_{8+\delta}$, as well as single crystals of $La_{2-x}Sr_xCuO_4$[19] (which involved a completely different surface preparation). We emphasize that our diamagnetic susceptibility measurement puts a bound of 1% on the superconducting fraction which seems too small to account for the signal to background ratio observed in Fig. 2b. The $T = 0$ superconductor-to-insulator transition is driven by quantum fluctuations[24] of the phase of the superconducting order parameter. The corresponding thermal fluctuations, which are vortex-like excitations in the pseudogap phase, have been probed by Nernst[25] and diamagnetism[26] experiments.

Our observations imply that the sharp quasiparticles of the d-wave superconducting state exist down to the lowest doping levels while rapidly loosing spectral weight, but are no longer visible on the insulating side. Nonetheless, a low energy d-wave like gap survives the phase-disordering transition. Obviously, the node must disappear as the Mott insulator is approached, as indicated by some photoemission studies[27]. We note that we are unable to make low temperature measurements at a smaller doping than that presented here due to sample charging, suggesting a fully gapped insulator.

Summarizing, we have found spectroscopic evidence for a d-wave nodal liquid ground state in the narrow doping regime between the high $T_c$ superconductor and the undoped Mott antiferromagnet. This quantum liquid has no superconducting order, the transport characteristics of an insulator, no sharp quasiparticles, and yet it has an energy gap that looks just like that of a *d*-wave superconductor. Since the spectral gap evolves smoothly through the insulator-to-superconductor phase transition, the *d*-wave superconductor appears to be just a phase-coherent version of the *d*-wave nodal liquid[28-30].

This work was supported by the U.S. NSF and the U.S. DOE, Office of Science. The Synchrotron Radiation Center is supported by the U. S. NSF.

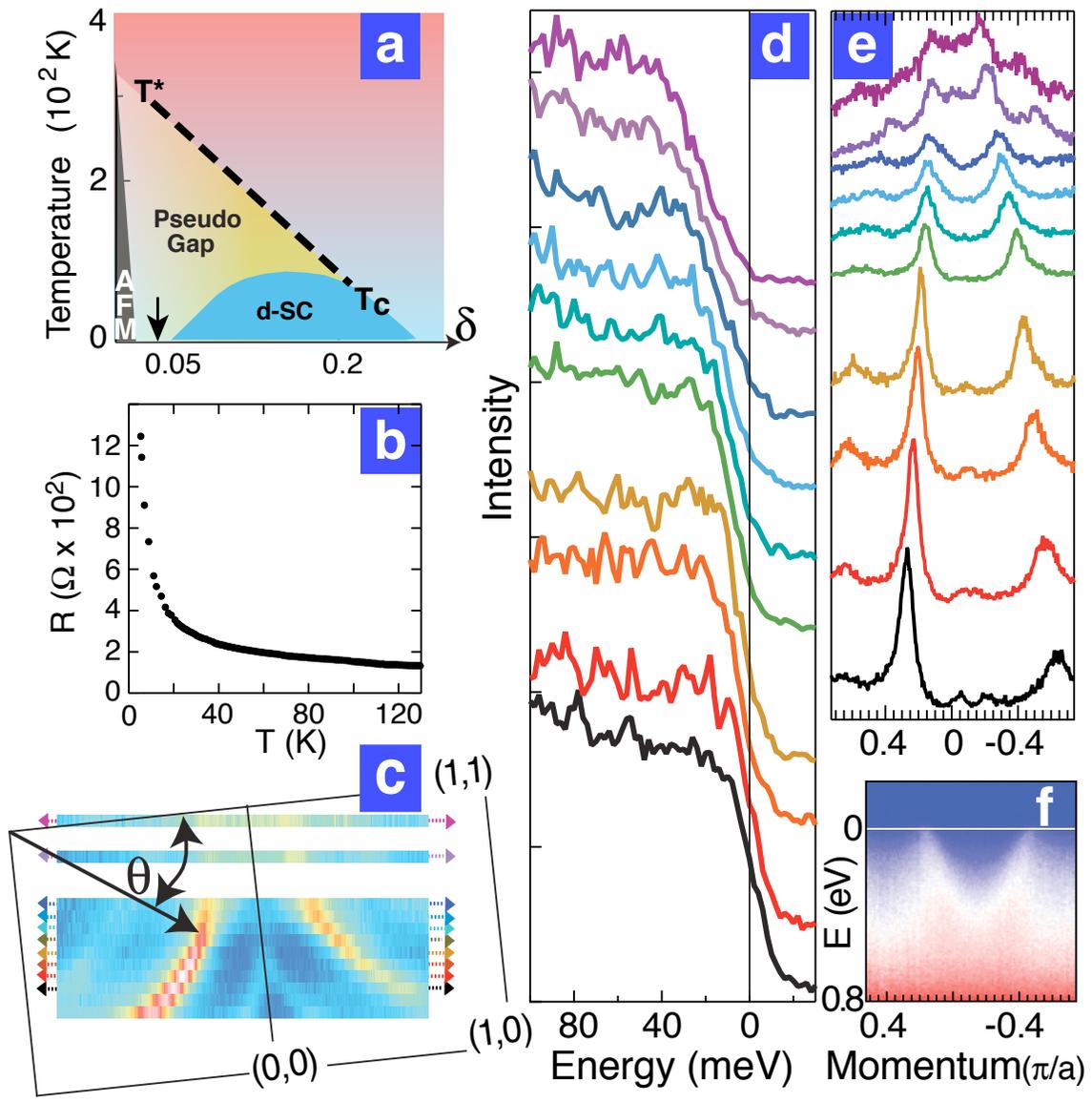

**Fig. 1**. **Data for non-superconducting samples.** (a) Schematic phase diagram of $Bi_2Sr_2CaCu_2O_{8+\delta}$ in the hole-doping ($\delta$), temperature ($T$) plane. The arrow indicates the doping level of the non-superconducting thin film samples whose data are shown in subsequent panels. (b) T-dependence of the resistance showing an insulating upturn. (c) Low-energy ARPES intensity at T=16K

(averaged over a 0 to 10 meV binding energy window) plotted as a function of **k** in the Brillouin zone. The high intensity points map out the underlying Fermi surface ($\mathbf{k}_F$) labeled by the angle $\theta$. (d) EDCs at various $\mathbf{k}_F$'s seen on the left-hand side of (c) with $\theta$ increasing from $0°$ (top) to $45°$ (bottom). (e) MDCs along the cuts marked in (c) using the same color coding in both panels. The top two MDCs are at binding energies of 18 and 22 meV respectively, all the rest correspond to zero energy. (f) Dispersion of the ARPES spectra for another sample, with the same doping as that shown in (b-e), along a momentum cut through the node.

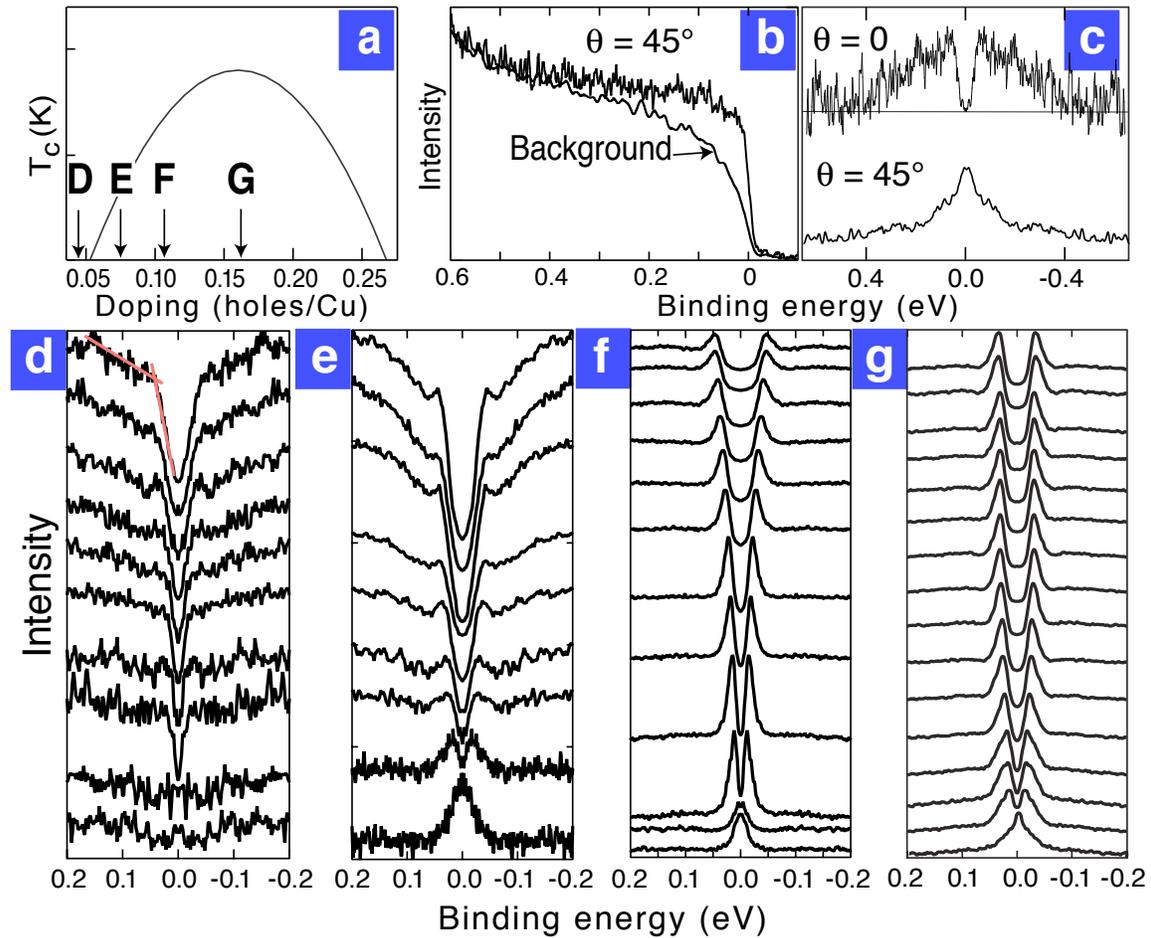

**Fig. 2. Spectral function vs. doping.** (a) The doping levels δ of the four samples whose spectra are shown in panels (d) through (g). The determination of δ is described in the Supplemental section. (b) Nodal spectrum and background (see text) for the non-superconducting sample D in (a) at $T = 16K$. **c** Background-subtracted, symmetrized intensity for sample D at the antinode ($\theta = 0°$) showing a gap and node ($\theta = 45°$) with zero gap. (d) through (g) show symmetrized EDCs, without background subtraction, for (d) the non-superconducting thin film at $T = 16K$, (e) an underdoped $T_c = 33K$ thin film at $T = 16K$; (f) an underdoped

$T_c$ = 69K single crystal at $T$ = 20K, and (g) a near-optimal $T_c$ = 80K thin film at $T$ = 40K. In each panel (d) through (g) the spectra are plotted with $\theta$ increasing from $0°$ (antinode) at the top to $45°$ (node) at the bottom. Note the highly anisotropic gap seen at all four doping levels, with sharp quasiparticle peaks at the gap edge at all $\mathbf{k}_F$, with weight diminishing with underdoping for all superconducting samples. Even after superconductivity is lost, we see in (d) a clear low-energy gap scale as emphasized by the red lines drawn in the top spectrum.

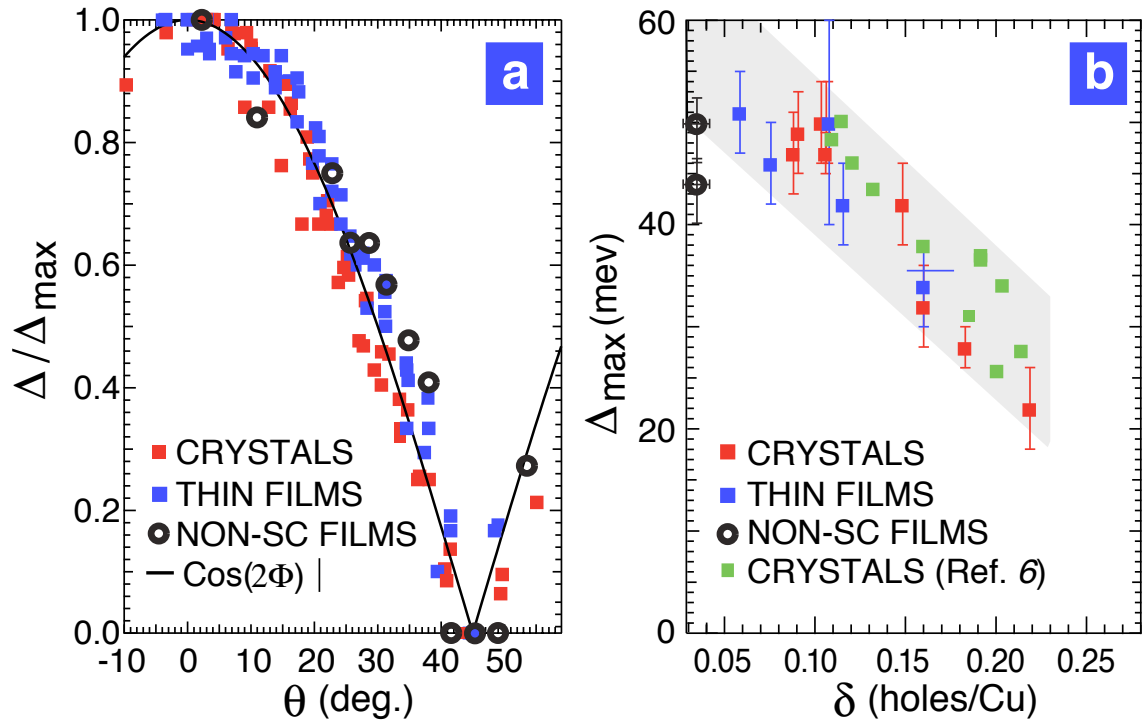

**Fig. 3**. **The spectral gap as a function of angle around the Fermi surface and doping.** (a) The spectral gap $\Delta(\theta)$, normalized by its maximum value at the antinode, plotted as a function of the Fermi surface angle $\theta$. Different colors are used for various superconducting samples (single crystals and thin films), while an open symbol is used for the non-superconducting sample. The energy gap at all doping levels is consistent with the d-wave form |cos($2\theta$)| shown as a black curve. (b) Maximum gap as a function of hole-doping. Gaps of the superconducting samples are denoted by filled symbols (blue for thin films and red for single crystals measured in this work, and green for published data[6]), while open circles are used for the non-superconducting samples.